# Information dynamics, natural computing and Maxwell's demon in two skyrmions system

Yoshishige Suzuki[1,2,3,*], Hiroki Mori[1,*], [†]Soma Miki[1,4,*], Kota Emoto[1,*], Ryo Ishikawa[5], Eiiti Tamura[1], Hikaru Nomura[1,**], Minori Goto[1,2,3,6]

1 Graduate School of Engineering Science, Osaka University, Toyonaka, 560-8531, Japan

2 Center for promotion of Advanced Interdisciplinary Research, Osaka University, Toyonaka, 560-8531, Japan

3 National Institute of Information Technology and Science, Advanced ICT research Instittute, Kobe, 651-2492, Japan

4 WPI Advanced Institute for Materials Research (AIMR), Tohoku University, Sendai, 980-8577, Japan

5 ULVAC-Osaka University Joint Research Laboratory for Future Technology, Osaka University, Suita, 565-0871, Japan

6 Graduate School of Engineering, University of Fukui, 3-9-1, Bunkyo, Fukui, Fukui 910-8507, Japan

* These authors are equaly contributed.

Present address

** International Center for Synchrotron Radiation Innovation Smart, Tohoku University, Sendai, 980-8579, Japan

†Corresponding Author: Soma Miki

WPI Advanced Institute for Materials Research (AIMR), Tohoku University, Sendai, 980-8577, Japan

Tel: +81-22-217-6004

e-mail: soma.miki.d8@tohoku.ac.jp

Abstract

The probabilistic information flow and natural computational capability of a system with two magnetic skyrmions at room temperature have been experimentally evaluated. Based on this evaluation, an all-solid-state built-in Maxwell's demon operating at room temperature is also proposed. Probabilistic behavior has gained attention for its potential to enable unconventional computing paradigms. However, information propagation and computation in such systems are more complex than in conventional computers, making their visualization essential.

In this study, a two-skyrmion system confined within a square potential well at thermal equilibrium was analyzed using information thermodynamics. Transfer entropy and the time derivative of mutual information were employed to investigate the information propagation speed, the absence of a Maxwell's demon in thermal equilibrium, and the system's non-Markovian properties. Furthermore, it was demonstrated that the system exhibits a small but finite computational capability for the nonlinear XOR operation, potentially linked to hidden information in the non-Markovian system. Based on these experiments and analyses, an all-solid-state built-in Maxwell's demon utilizing the two-skyrmion system and operating at room temperature is proposed.

How can we design a computing machine that operates with minimal energy[1]? To address this question, Bennett proposed the concept of Brownian computation[2], in which thermal energy drives the exploration process, and the computation proceeds once the correct solution is identified. To further develop practical implementations of such circuits, Peper introduced a universal Brownian computation circuit based on token dynamics[3]. In this approach, a mechanism analogous to Maxwell's demon, called a Conservation-Join, is employed to determine the correct answer. Potential candidates for tokens include electrical charges in single-electron transistors and skyrmions in magnetic thin films. However, single-electron transistors can only function at extemely low temperatures [4], limiting their paractical applications.

Magnetic skyrmions are particle-like magnetic structures that emerge in continuous magnetic ultrathin films and are expected to find applications in various memory and computing devices, such as race track memory [5] and Brownian computers [6]. Race track memory utilizes the translational motion of skyrmions driven by electric current [7] and their non-volatility. In contrast, Brownian computers leverage the probabilistic behavior resulting from the Brownian motion of skyrmions at room temperature[8-12] and their mutual repulsive interactions[13,14], enabling intelligent computations distinct from conventional computers. To realize the Brownian computer with ultra-low energy consumption, the techniques for potential control using film structure and/or ion irradiation [6, 15] and for electric field manipulation[16-18] have been developed. So far, applications to brain-like computers, including Brownian computers [2], reservoir computing[19-22], and cellular automata [23], have been proposed, yielding preliminary results.

In these computing systems, information propagates in a complex manner, making it impossible to understand using the simple information propagation models employed in conventional computers. Therefore, it is critically important to visualize how information propagates and to elucidate the computational capabilities available. Transfer entropy[24] is a useful physical quantity to clarify the information dynamics in complex systems. Transfer entropy quantifies the causal flow of information between time series data from the probability distribution of a system[25]. It is effective in systems that handle complex time-series data, and has wide-ranging applications such as biology[24], neuroscience[26, 27], physics[28], economics[29], food web[30], and SNS[31].

This paper experimentally investigates the information dynamics of thermal motion when two skyrmions are confined within a square region. Skyrmions exchange information through their repulsive interactions. The flow of information in thermal equilibrium was analyzed using transfer entropy and the subsystem-timederivative of mutual information[32]. The latter allows us to determine the existence/non-existence of Maxwell's demon in this physical system. Furthermore, the non-Markovian nature of the system and its capability for natural computation were also examined. Finally, we propose a setup to realize an autonomous Maxwell's demon in a skyrmion system and demonstrate its operation through analytical calculations.

The sample structure is shown in Figure 1(a). Ta (5.0 nm)|$Co_{16}Fe_{64}B_{20}$ (1.3 nm)|Ta (0.24 nm)|MgO (1.5 nm)|$SiO_2$ (3.0) was deposited onto the Si|$SiO_2$ substrate using magnetron sputtering system (Canon ANELVA, E-880SM in Osaka University). By additionally depositing 0.11 nm of $SiO_2$ on a portion of the surface of this film, skyrmions can be confined within that area[6, 23]. We used photolithography to confine two skyrmions in an 8 μm × 8 μm square cell (Figure 1 (b)). Figure 1(c) is a MOKE microscope image of the sample, and the area surrounded by the red line represents the area where $SiO_2$ was additionally deposited. Two skyrmions with about 1.5 μm in diameter are confined. Brownian motion under two skyrmions interacting with each other was observed.

Observations were taken in video for 10 seconds at a frame rate of 250 fps (time step $\Delta t$=4 ms). The number of records was taken 75 times (in total 187,500 frames). During observation, a magnetic field was applied perpendicularly upward to the sample surface, and the sample temperature was controlled using a heater. The optimal conditions were at around 0.25 mT and around 51 °C. At elevated temperature, skyrmions show a high diffusion constant as high as around 100 μm²/s (see suplimental information) and 34.4 μm²/s at 51.5 °C.

In order to investigate information dynamics in the skyrmion system, we created time series data from the trajectory of skyrmions exhibiting Brownian motion inside the square cell. Figure 1(d) shows the trajectory of each skyrmion. The one with the larger $y$-coordinate of the initial position is defined as skyrmion A, and the one with the smaller initial $y$-coordinate is defined as skyrmion B. The solid line shown in yellow represents the trajectory of Skyrmion A, and the solid line shown in red represents the trajectory of Skyrmion B.

Here, $A_n$ and $B_n$ are the random variable of state (position) of skyrmion A and B at discrete time $n$, respectively. We consider the observed $x$-position (event) of $a_n$ and $b_n$ of skyrmions A and B, which are two time-series data sets $\{a_1, a_2, ..., a_n, ..., a_N\}$ and $\{b_1, b_2, ..., b_n, ..., b_N\}$, respectively. In our experiments, the time step is equal to 4 ms, and the position data are binarized. We set a threshold value for the $x$-coordinate of each skyrmion and binarized it. Coordinates smaller than the threshold is defined as state "0", and coordinates larger than the threshold is defined as state "1". The threshold value is determined so that the number of cases of "0" and "1" would be equal for all position coordinates of Skyrmion A (B). In this analysis, this binary probability distribution was used.

Generally, the transfer entropy from skyrmion $A$ to $B$ between discrete time $n$ and $n$+1 is defined by equation (1)[25].

$$I_{A\to B}^{tr} = S\big(B_{n+1}\big|B_1, ..., B_n\big) - S\big(B_{n+1}\big|A_1, ..., A_n, B_1, ..., B_n\big) \tag{1}$$

Here $S(B_{n+1}|A_1, ..., A_n, B_1, ..., B_n)$ is the conditional Shannon entropy[33, 34]. The Shannon entropy $S(B_{n+1})$ of the random variable of $B$ at discrete time $n$+1 is expressed by $S(B_{n+1}) = -k_B \sum_{b_{n+1}} p(b_{n+1}) \ln p(b_{n+1})$, where $p(b_{n+1})$ is the probability of finding skyrmion B in position $b_{n+1}$. Conditional Shannon entropy is the Shannon entropy for conditional probabilities, i.e., $S(B_{n+1}|B_1, B_2, ..., B_n) \equiv -k_B \sum_{b_1, b_2, ..., b_{n+1}} p(b_1, b_2, ..., b_{n+1}) \ln p(b_{n+1}|b_1, b_2, ..., b_n)$. The transfer entropy is a reduction in skyrmion B's conditional entropy by knowing the past positions of skyrmion A and expressing a causal information flow from skyrmion A to B. To discuss information dynamics, we extend the above definition to longer time separation between cause and result as follows;

$$\begin{cases} I_{A\to B}^{TE}(j) = S\big(B_{n+j}\big|B_n\big) - S\big(B_{n+1}\big|A_n, B_n\big) \\ I_{B\to B}^{TE}(j) = S\big(B_{n+j}\big|A_n\big) - S\big(B_{n+1}\big|A_n, B_n\big) \end{cases} \tag{2}$$

Here, time separation is changed from 1-time step to $j$-time steps. In addition, information from the past, i.e., $n$-1,…1, are neglected. Later treatment facilitates numerical analysis using a limited number of data sets and does not affect the results if the system is Markovian. The first line uses the extended definition of the transfer entropy (TE) from A to B. The transfer entropy from B to B is also defined for comparison purposes in the second line.

Figure 2(a) shows the schematic diagram (3-node diagram) of the transfer entropy $I_{\text{A}\to\text{B}}^{\text{TE}}(j)$ from the state of the current skyrmion $A_n$ to the state of future skyrmion $B_{n+j}$ in a situation where the state of current skyrmion $B_n$ is known. This evaluates the transfer entropy flowing between two different skyrmions. In the range where $j$ is large, the correlation between $B_{n+j}$ and $(A_n, B_n)$ reduces, thus $I_{\text{A}\to\text{B}}^{\text{TE}}(j)$ is expected to approach 0. Figure 2(b) is the schematic diagram of the transfer entropy $I_{\text{B}\to\text{B}}^{\text{TE}}(j)$ from the state of the current skyrmion $B_n$ to the state of future skyrmion $B_{n+j}$ in a situation where the state of current skyrmion $A_n$ is known. This represents the transfer entropy of same skyrmion from its own past to the future.

Figure 2(c) shows the time-separation dependence of the transfer entropies in the two skyrmion systems evaluated from experimental data using three-node analysis (equation (2)). The vertical axis of the figure is the value of transfer entropy normalized by 1-bit Shannon entropy. The larger the value, the more information is flowing, and the state is completely determined when the value is 1.0. The horizontal axis represents time separation, and the larger $j$ indicates the transfer of entropy to the distant future. Plotting was performed at 38 points from $j$=0 to 37, and was converted to the actual time separation. Since the transfer entropy $I_{\text{A}\to\text{B}}^{\text{TE}}(j)$ is not defined for $j$=0, it is not evaluated. Results represented by pink dots indicate $I_{\text{A}\to\text{B}}^{\text{TE}}(j)$, and results represented by blue dots indicate $I_{\text{B}\to\text{B}}^{\text{TE}}(j)$. The result represented by pink and blue opened dots are the transfer entropies evaluated from two random number sequences of 0 and 1 and correspond to the statistical error of the analysis in this study. In a range with values comparable to this statistical error, the results cannot be determined to be significant. Since both $I_{\text{A}\to\text{B}}^{\text{TE}}(j)$ and $I_{\text{B}\to\text{B}}^{\text{TE}}(j)$ have values larger than the statistical error of the analysis in the range where the time separation is less than 0.1 seconds, it can be seen that significant results were obtained.

$I_{\text{A}\to\text{B}}^{\text{TE}}(j)$ increases in value from $j$=1 and reaches its maximum value at the time separation between $j$=2 and 3 (approximately 10 ms seconds of the time separation. See the inset of Figure 2(c).). This indicates that skyrmion B receives the largest amount of information from skyrmion A at this time. Afterward, it attenuates as $j$ increases. The diffusion length of the skyrmion over 10 ms is about 0.8 μm. It is close to the skyrmion radius (0.8 μm). Therefore, the peak structure in the transfer entropy graph is attributed to the fact that information transfer requires a transient transition of the skyrmion B's position from 0 to 1 or from 1 to 0, necessitating a finite time corresponding to the diffusion time needed to switch the skyrmion position[35]. On the other hand, $I_{\text{B}\to\text{B}}^{\text{TE}}(j)$ is equal to 1 at $j$=0, decreases rapidly at first, and then decays exponentially with a smaller slope as a function of $j$.

In addition, the attenuation time $\tau$ was evaluated from the fitting of the logarithmic attenuation region in the graphs, using the fitting function of $\alpha \exp[-j / f\tau]$. Here, $f$ is the frame rate of 250 fps and $\alpha$ is a coefficient. As a result, it was found that the attenuation time of $I_{A \to B}^{TE}(j)$ is about 21 ms, and that of $I_{B \to B}^{TE}(j)$ is about 26 ms, indicating that skyrmions retain propagated information for more than 20 ms. The initial rapid decrease in $I_{B \to B}^{TE}(j)$ is thought to correspond to the dissipation of information during the diffusion process of a single skyrmion. On the other hand, the subsequent decrease in transfer entropies, which occurs at nearly the same ratet for both entropies but with a longer decay time, suggests that this process corresponds to the slower dissipation of information due to the sharing of information between two skyrmions.

To validate the Markov approximation, we performed an analysis of transfer entropy considering the contribution of the previous node using the following equations (3)

$$\begin{cases} I_{A \to B}^{TE(2)}(j) = S\big(B_{n+j} \big| B_n, B_{n-1}\big) - S\big(B_{n+1} \big| A_n, A_{n-1}, B_n, B_{n-1}\big) \\ I_{B \to B}^{TE(2)}(j) = S\big(B_{n+j} \big| A_n, A_{n-1}\big) - S\big(B_{n+1} \big| A_n, A_{n-1}, B_n, B_{n-1}\big) \end{cases} \tag{3}$$

The results are shown with plus signs in Fig. 4(c). There is a slight difference between the analysis assuming a Markov process and the analysis considering the contribution of the previous node, but the general time dependence remains unchanged. The statistical errors for the analysis, considering the previous node, are indicated by crosses. It is observed that the statistical errors increase as the number of nodes is increased.

It was pointed out that the subsystem-time derivative of the mutual entropy expresses a directional information flow (IF) (4) [32].

$$\dot{I}_{A \to B}^{IF} \equiv \lim_{\Delta t \to 0} \big(I_{A \leftrightarrow B}^{Mutual}(1) - I_{A \leftrightarrow B}^{Mutual}(0)\big) / \Delta t \tag{4}$$

In the above expression, the mutual information $I_{B \leftrightarrow A}^{Mutual}(j)$ is defined by 1st line of the following expressions.

$$\begin{cases} I_{A \leftrightarrow B}^{Mutual}(j) \equiv S\big(B_{n+j}\big) - S\big(B_{n+j} \big| A_n\big) \\ I_{B \leftrightarrow B}^{Mutual}(j) \equiv S\big(B_{n+j}\big) - S\big(B_{n+j} \big| B_n\big) \end{cases} \tag{5}$$

As in the above definition, the mutual information in the first and second lines are equivalent to the non-local and local active information storages, respectively [36]. The above mutual information (active information storage) are plotted in Figure 3 to observe the time derivative as a slope of a graph.

In Figure 3 (a) and (b), the schematic diagram (2-node diagram) of the above-defined mutual entropy is shown. Figure 3(c) shows the time-separation dependence of the mutual information evaluated from experimental data using two-node analysis (equation(5)). The vertical axis of the figure is the value of the mutual information. The horizontal axis is the time separation, $j$. Plotting was performed at 38 points from $j$=0 to 37. Results represented by pink dots indicate $I_{A \leftrightarrow B}^{Mutual}(j)$, and results represented by blue dots indicate $I_{B \leftrightarrow B}^{Mutual}(j)$. The closed and opened dots in light blue indicate statistical error. It can be seen below 0.12 seconds significant results were obtained.

Both mutual information are monotonically decreasing as the time separation increases. At the beginning ($j$=0 to 1), however, the behaviors are very different between two graphs. The slope of $I_{A \leftrightarrow B}^{Mutual}(j)$ is almost zero at $j$=0, and that of $I_{B \leftrightarrow B}^{Mutual}(j)$ shows the largest slope there. The initial slope of the $I_{A \leftrightarrow B}^{Mutual}(j)$ corresponds to $\dot{I}_{A \rightarrow B}^{IF}$. Here, it should be mentioned that $I_{B \leftrightarrow B}^{Mutual}(j)$ is almost equivaent to $I_{B \rightarrow B}^{TE}(j)$. This means that information flows from A to B is minor compared with B to B flow. It is clear from the fact that $I_{A \rightarrow B}^{TE}(j)$ is about 1/15 of $I_{B \rightarrow B}^{TE}(j)$.

In a steady state, the total time derivative of the mutual information is zero. Therefore, $\dot{I}_{A \rightarrow B}^{IF} = -\dot{I}_{B \rightarrow A}^{IF}$. This means that if $\dot{I}_{A \rightarrow B}^{IF} > 0$, Skyrmion B acquires information about Skyrmion A, thereby imparting negative entropy to Skyrmion A. And, Skyrmion B can be regarded as a Maxwell's demon for $\dot{I}_{A \rightarrow B}^{IF} > 0$. In our experiment, the information flow between subsystems was zero, and no Maxwell's demon was observed. This is a natural and reasonable result for a system in thermal equilibrium.

Figure 4 shows the results of model calculations based on the master equation, assuming a bipartite Markov jump process [32]. The parameters used are the interaction energy between skyrmions ($\varepsilon_I$) and the jumping rate ($R_0$) (see Supplementary Information). By setting $\varepsilon_I$ =0.32 $k_B T$ and $R_0$=19.9 [s$^{-1}$], the theoretical curve closely fits the graph of the mutual information between skyrmion A and future skyrmion B. On the other hand, the theoretical model fails to explain the initial rapid decay of mutual information between the present and future skyrmion B (self-active information storage). This discrepancy is attributed to the reduction of hidden information associated with the trap sites of the skyrmions (discussed later). This hidden information is related to non-Markovian properties and is not incorporated into the Markov jump model adopted here.

As shown in Fig. 2(c), the analysis results of transfer entropy, assuming a Markov process, were slightly modified by taking into account the contribution of the preceding nodes. This indicates that the system comprising the two skyrmions under consideration exhibits a non-Markovian response. To investigate the non-Markovian property of our system, we compared the following two mutual information.

$$\begin{cases} I_{Present \leftrightarrow Future}^{Mutual}(j) = S\left(B_{n+j} \middle| A_{n-1}, B_{n-1}\right) - S\left(B_{n+j} \middle| A_{n-1}, B_{n-1}, A_n, B_n\right) \\ I_{Past \leftrightarrow Future}^{Mutual}(j) = S\left(B_{n+j} \middle| A_n, B_n\right) - S\left(B_{n+j} \middle| A_n, B_n, A_{n-1}, B_{n-1}\right) \end{cases} \quad (5)$$

The first equation, $I_{Present \leftrightarrow Future}^{Mutual}(j)$ represents the mutual information between the current skyrmion $\left(B_n, A_n\right)$ and the future skyrmion $\left(B_{n+j}\right)$, given the past position $\left(B_{n-1}, A_{n-1}\right)$. Since the system continually generates new information, this value is expected to be finite. The second equation, $I_{\text{Past}\leftrightarrow\text{Future}}^{\text{Mutual}}(j)$ represents the mutual information between the past skyrmion $(B_{n-1}, A_{n-1})$ and the future skyrmion $(B_{n+j})$, given the current position

$(B_n, A_n)$. If the system is Markovian, this value should be zero. In Figure 5(a) and (b), diagrammatic expressions

(3-nodes) of $I_{\text{Present}\leftrightarrow\text{Future}}^{\text{Mutual}}(j)$ and $I_{\text{Past}\leftrightarrow\text{Future}}^{\text{Mutual}}(j)$ are shown, respectively.

In Figure 5(c), the experimentally obtained values of $I_{\text{Present}\leftrightarrow\text{Future}}^{\text{Mutual}}(j)$ and $I_{\text{Past}\leftrightarrow\text{Future}}^{\text{Mutual}}(j)$ are shown as red and green dots, respectively. Light blue closed and open dots represent statistical errors. As expected,

$I_{\text{Present}\leftrightarrow\text{Future}}^{\text{Mutual}}(j)$ exhibits a finite value. Notably, $I_{\text{Past}\leftrightarrow\text{Future}}^{\text{Mutual}}(j)$ also has a finite value. This result indicates that the system of our two skyrmions is non-Markovian. However, the information flow from the past is small, with the ratio of $I_{\text{Present}\leftrightarrow\text{Future}}^{\text{Mutual}}(j)$ to $I_{\text{Past}\leftrightarrow\text{Future}}^{\text{Mutual}}(j)$ being approximately 6.5 at $j$=1. This suggests that the transfer entropy, which ignores the past as defined in Equation (2), is approximately valid.

The reason why the skyrmion system in this study exhibits non-Markovian behavior is believed to lie in
the coarse-graining (binarization) of the skyrmion positions. Figure 6 presents a heat map showing the spatial
distribution of the integrated-skyrmion's-staying-time. The brightness is proportional to the logarithm of the
staying time at each pixel. Considering that the staying time in thermal equilibrium follows an exponential
relationship with the potential depth, the brightness reflects the depth of the skyrmion's trapping potential.
Approximately 15 to 20 trapping potentials can be observed in the figure. This means that each binary value
corresponds to about 7 to 10 trapping positions, and in the binarized data sequence used for analysis, information
about which specific trap the skyrmion is captured in is lost. Systems with such hidden degrees of dynamical
freedom generally exhibit non-Markovian behavior. Furthermore, in the field of machine learning, such dynamical
systems are referred to as hidden Markov systems, which are known to possess computational capability and are
applied to tasks such as filtering and maximum likelihood estimation[37].

The coupling between the hidden degrees of freedom in the skyrmion system studied in this paper was
not designed for computational purposes, but it may still be performing some form of computation. To investigate
this, we checked whether XOR computation is being performed by the skyrmions. XOR computation, as
expressed by Equation (6), is a nonlinear operation that requires a certain level of computational capability.

$$x_n = \frac{1}{2} - 2\left(a_n - \frac{1}{2}\right)\left(b_n - \frac{1}{2}\right) \tag{6}$$

Figure 7(a) shows a diagram of the evaluated mutual information. First, we performed an XOR
computation on the positional data of skyrmions A and B at the current time, treating the result as the first node.
Then, we calculated the mutual information between this node and the future position of skyrmion B. If the
mutual information is $k_{\text{B}} \ln 2$, it implies that the result of the XOR computation on the current positions of
skyrmions A and B can be predicted with 100% accuracy from the future position of skyrmion B. This would indicate the presence of XOR computational capability.

$$I_{\text{XOR}\leftrightarrow\text{B}}^{\text{Mutual}}(j) = S(X_n) - S(X_n|B_{n+j}) \qquad (7)$$

Figure 7(b) shows the time interval dependence of the mutual information as orange dots, as described in Equation (7). Light blue dots represent statistical errors. It can be seen that significant results are obtained for time separation below 0.04 seconds. From the value of mutual information, it is evident that the skyrmion system performs XOR computation at a rate of approximately 0.04% for $j$=1. Although the value is small, it is statistically significant, and the fact that a very simple two-skyrmion system performs nonlinear natural computation in thermal equilibrium is intriguing.

Figure 4(b) presents a conceptual diagram of a circuit designed to implement Maxwell's demon in a skyrmion system. For simplicity, the skyrmion film is modeled as a channel in which skyrmions A and B can move only in the horizontal direction. Additionally, magnetic tunnel junction (MTJs) is constructed at position 0 of skyrmion A, enabling electrical detection of the skyrmions. An electrode is installed at position 0 of skyrmion B via a lower resistance barrier. When the MTJs are turned on, a voltage is applied to the electrode to prevent skyrmion B from entering the region (see Supplementary Information for details on the realistic circuit configuration). The control can be done by voltage-controlled magnetic anisotropy (VCMA) [8, 17, 18]. Through this construction, skyrmion B can obtain position information of skyrmion A.

As a result, the mutual information between skyrmion A and future skyrmion B (pink dashed line in Figure 4(a) exhibits a positive slope near zero time lag, indicating the presence of an information flow from skyrmion A to skyrmion B. The mutual information was obtained numericaly solving the master equation, which includes detection by MTJ and voltage application. The only newly introduced paprameter for the calculation is the potential rise, $\varepsilon_v$=1 $k_\text{B}T$, when voltage is applied. The information flow from A to B confirms that skyrmion B functions as Maxwell's demon. As shown in the graph, the information flow from skyrmion B to skyrmion A is negative (green dashed line), suggesting that skyrmion A undergoes cooling or enables the extraction of work.

Under the influence of Maxwell's demon, detailed balance is valid only within each local system-i.e., the skyrmion A system and the skyrmion B system. Consequently, the total system is not in thermal equibrium but rather in a steady state. In this situation thermodynamic flows are induced within the system. Figure 4(c) illustrates the dependence of various entropy-related flows on the strength of the skyrmion interaction. The information flow into B (pink) is slightly smaller than the heat emission (red) but slightly larger than the absolute value of the cooling effect (light blue). This suggests that the skyrmion system has the potential to form a built-in Maxwell demon without requiring an external circuit, utilizing tunnel magnetoresistance elements and the VCMA effect. Furthermore, this circuit is expected to operate at much higher temperatures than a single-electron transistor. In this study, we used relatively large skyrmions (1.5 μm) to enable observation with an optical microscope. However, skyrmions can be scaled down to approximately 10 nm, which would allow for an expected operation speed of around 1 nano-second. Additionally, in smaller skyrmions, a stronger repulsive force arises between skyrmions due to exchange interactions, making room-temperature operation feasible.

In this study, we conducted an information thermodynamic analysis of the thermal equilibrium state of two relatively large skyrmions confined within a box-shaped potential. Due to the repulsive force between the skyrmions, information transfer occurs. Analyzing transfer entropy revealed that information transmission between the skyrmions takes approximately 10 milliseconds.

An analysis of information flow, obtained from the time derivative of mutual information, confirmed that no information feedback mechanism resembling Maxwell's demon exists in this system. Futthermore, an analysis of the system's Markovian properties demonstrated that it exhibits non-Markovian behavior. Additionally, it was confirmed that the system can performs XOR computations, albeit at a minimal level. This can be regarded as a form of natural computation driven by thermal equilibrium fluctuations. Moreover, model calculations indicated that a built-in Maxwell's demon could be realized through detection via the magnetoresistance effect and control via voltage modulation within the device.

Skyrmions represent a rare system in which Brownian motion occurs in a solid at room temperature without involving material transport. As a model for information thermodynamic systems at room temperature, skyrmions hold potential for future applications, such as the development of autonomous Maxwell's demons and other information thermodynamic engines, including ultra-low power computation circuits.

Acknowledgement

This research was supported by JSPS Grant-in-Aid for Scientific Research(S) Grant Numbers JP20H05666, JST CREST Grant Numbers JPMJCR20C1, and Grant-in-Aid for Research Activity Start-up Grant Numbers 24K22860.

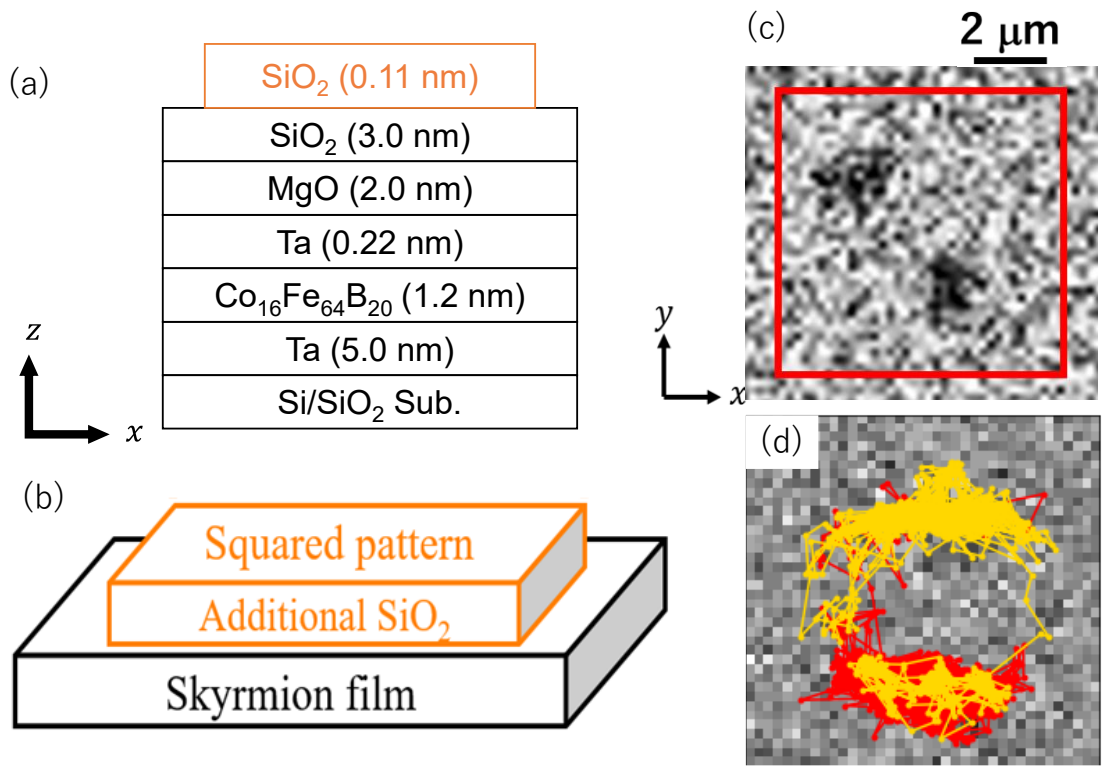

Figure 1 (a) Schematic diagram of the sample structure. Skyrmions are formed in the $Co_{16}Fe_{64}B_{20}$ layer (atomic percentages are indicated in subscript). (b) By depositing an additional $SiO_2$ layer (shown in orange) on top of the existing $SiO_2$, skyrmions are confined within that region, forming a potential box. (c) MOKE microscope image of skyrmions confined within the box. The red line represents the estimated edge of the box, determined based on skyrmion behavior. (d) The yellow and red solid lines indicate the trajectories of skyrmions A and B, respectively.

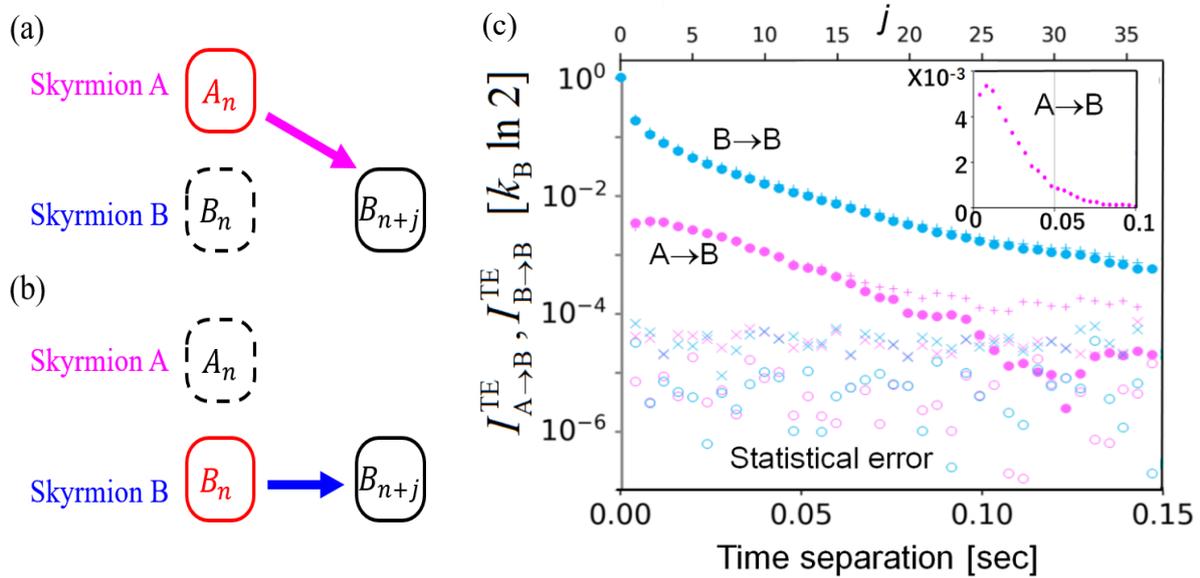

Figure 2 (a) Schematic diagram of the transfer entropy $I_{A \to B}^{TE}(j)$ from the current skyrmion A to the future skyrmion B, given that the current skyrmion B is known. The nodes shown as dotted lines represent already known nodes, while the nodes shown in red represent newly learned nodes. (b) Schematic diagram of the transfer entropy $I_{B \to B}^{TE}(j)$ from the current skyrmion B to the future skyrmion B, given that the current skyrmion A is known. (c) Experimentally obtained transfer entropies as functions of time separation. The results shown as pink and blue closed circles represent $I_{A \to B}^{TE}(j)$ and $I_{B \to B}^{TE}(j)$, respectively. Open circles indicate the statistical errors of the analysis. Pink and blue plus signs (+) represent the transfer entropy considering five nodes (see Eq. (3)), while cross marks (×) indicate the corresponding statistical errors.

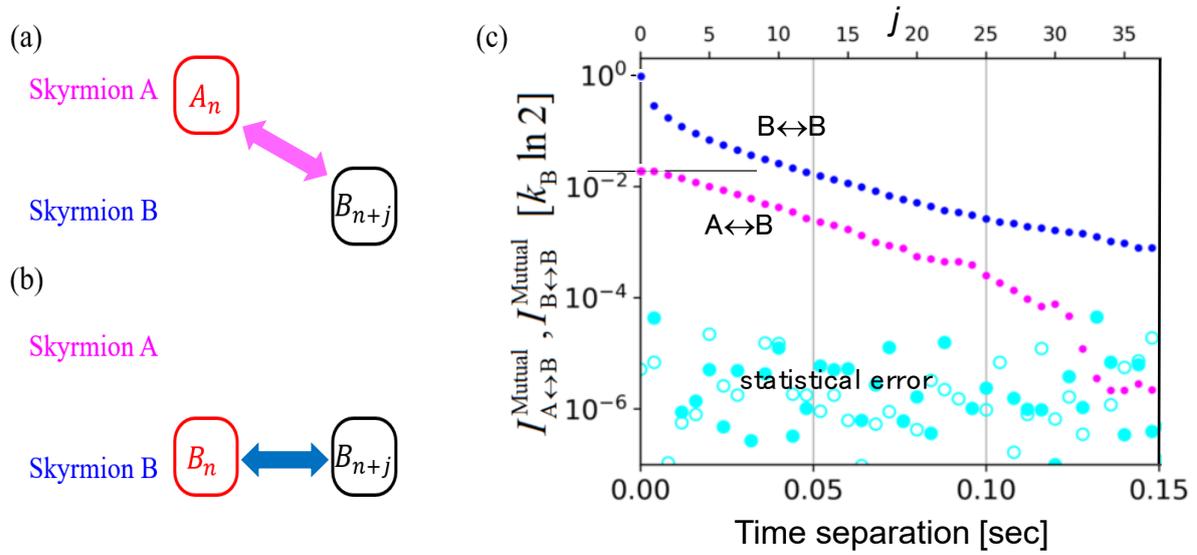

Figure 3 (a) Schematic diagram of the mutual information $I_{\text{B}\leftrightarrow\text{A}}^{\text{Mutual}}(j)$ between the current skyrmion A and the future skyrmion B. (b) Schematic diagram of the mutual information $I_{\text{B}\leftrightarrow\text{B}}^{\text{Mutual}}(j)$ between the current skyrmion B and the future skyrmion B. (c) Experimentally obtained mutual entropies as functions of the time separation. Results shown in pink and blue indicate $I_{\text{B}\leftrightarrow\text{A}}^{\text{Mutual}}(j)$ and $I_{\text{B}\leftrightarrow\text{B}}^{\text{Mutual}}(j)$, respectively. Results shown in light blue represent the statistical error of the analysis.

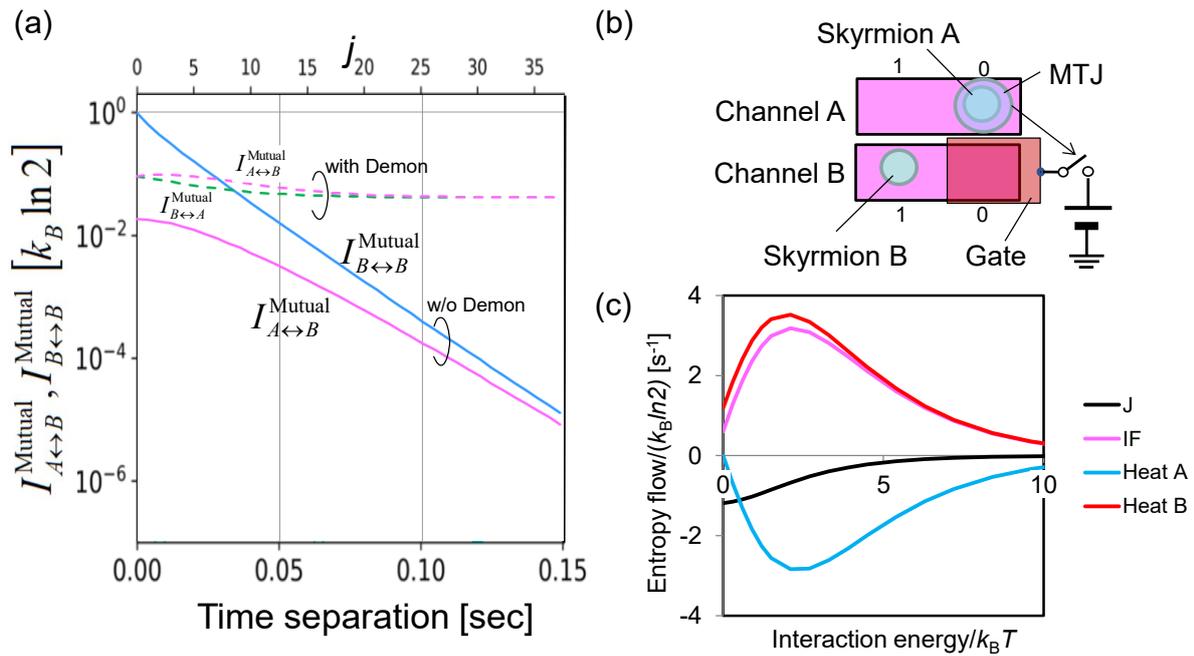

Figure 4. Model calculations based on a bipartite Markov jump process. (a) Dependence of mutual information on the time separation between the present and future skyrmion B in the absence of Maxwell's demon (solid blue line). Mutual information between the present skyrmion A and the future skyrmion B is shown as a solid pink line. In the presence of Maxwell's demon, the mutual information between the present skyrmion A (B) and the future skyrmion B (A) is represented by pink (green) dashed lines. (b) Conceptual diagram of a skyrmion circuit that implements Maxwell's demon. When the magnetic tunnel junction (MTJ) is turned on by the presence of a skyrmion, a voltage is applied to the red electrode, preventing skyrmion A from entering this location. (c) Dependence of various entropy-related flows on skyrmion repulsion energy in the presence of Maxwell's demon. The probability flow (J: black), information flow from A to B (IF: pink), heat emission from B (red), and heat emission from A (light blue) are shown.

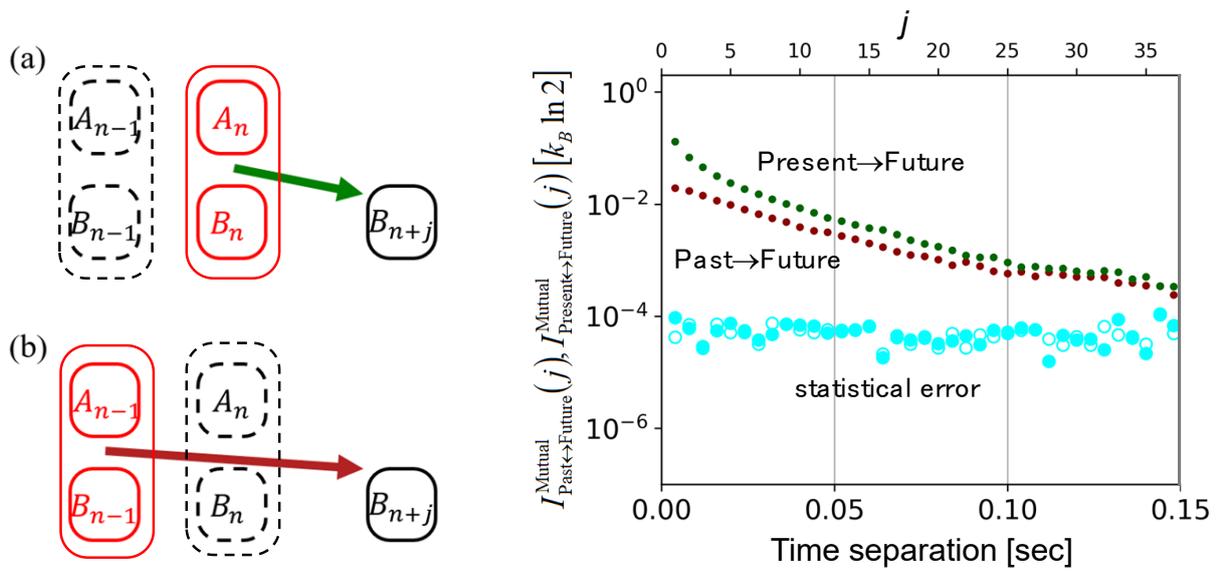

Figure 5 (a) Schematic diagram of the mutual information $I_{\text{Present}\leftrightarrow\text{Future}}^{\text{Mutual}}(j)$ between $(A_n, B_n)$ and $B_{n+j}$ when the past configuration $(A_{n-1}, B_{n-1})$ is known. (b) Schematic diagram of the mutual information $I_{\text{Past}\leftrightarrow\text{Future}}^{\text{Mutual}}(j)$ between $(A_{n-1}, B_{n-1})$ and $B_{n+j}$ when the present configuration $(A_n, B_n)$ is known. (c) Experimentally obtained mutual information as function of the time separation. Dark green and red dots represent $I_{\text{Present}\leftrightarrow\text{Future}}^{\text{Mutual}}(j)$ and $I_{\text{Past}\leftrightarrow\text{Future}}^{\text{Mutual}}(j)$, respectively. Light blue points represent the statistical error of the analysis.

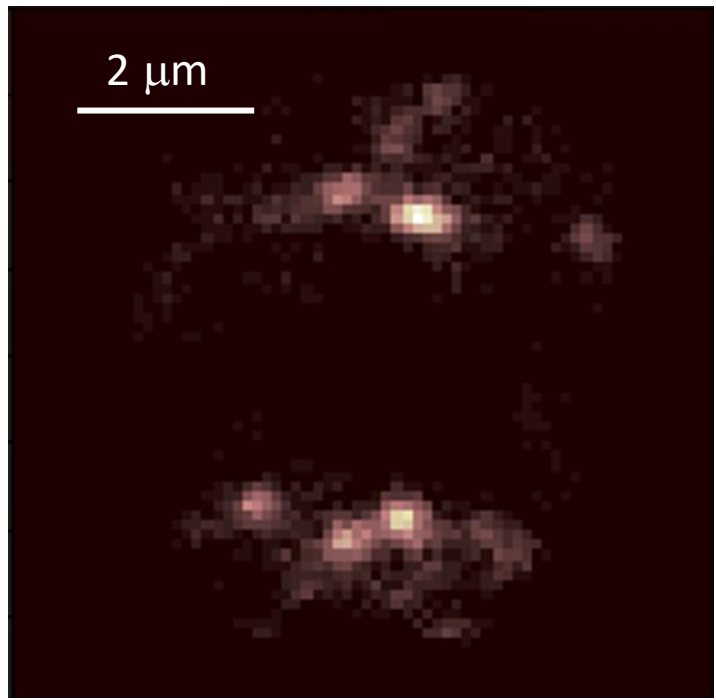

Figure 6 Trapping potential distribution within the potential box. Brighter color indicates deeper potential energy. The potential was estimated based on the skyrmion's residence time distribution.

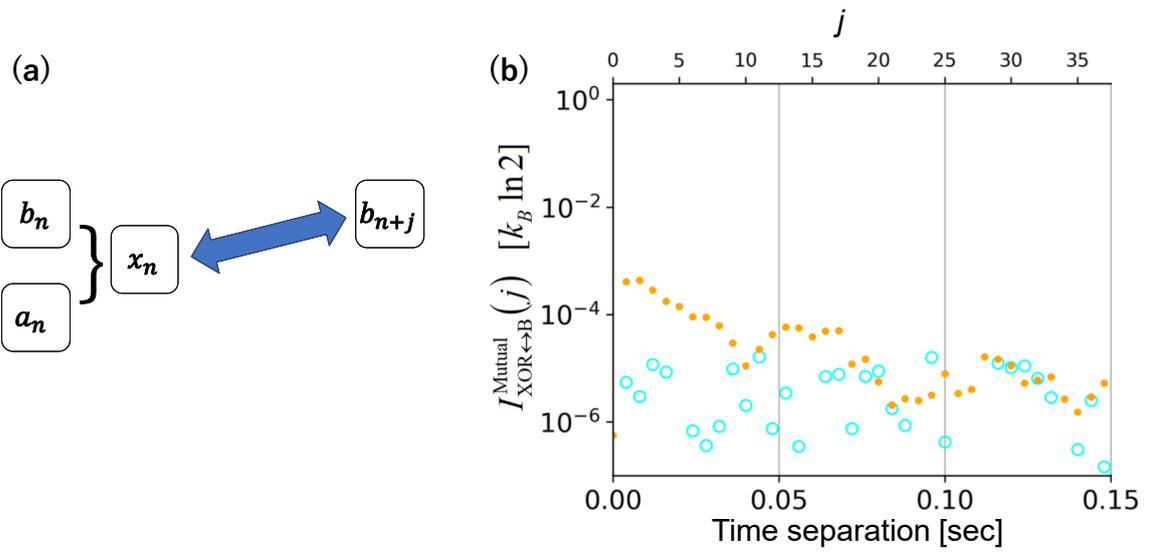

Figure 7 (a) Schematic diagram of the mutual information $I(X_n : B_{n+j})$ where $X_n$ represents the XOR operation between $B_n$ and $A_n$. (b) Mutual information $I(X_n : B_{n+j})$ (shown in orange). Light blue markers indicate the statistical error of the analysis.